\documentclass[prd,aps,twocolumn,nofootinbib,preprintnumbers]{revtex4}
\usepackage{graphicx,amsmath}
\usepackage{comment}
\usepackage{color}

\begin{document}


\newcommand{\mtrem}[1]{{\color{red} \bf $[[ $ MT: #1$ ]]$}}

\def\beq{\begin{eqnarray}}
\def\eeq{\end{eqnarray}}
\newcommand{\gsim}{ \mathop{}_{\textstyle \sim}^{\textstyle >} }
\newcommand{\lsim}{ \mathop{}_{\textstyle \sim}^{\textstyle <} }
\newcommand{\vev}[1]{ \left\langle {#1} \right\rangle }
\newcommand{\bra}[1]{ \langle {#1} | }
\newcommand{\ket}[1]{ | {#1} \rangle }
\newcommand{\EV}{ {\rm eV} }
\newcommand{\KEV}{ {\rm keV} }
\newcommand{\MEV}{ {\rm MeV} }
\newcommand{\GEV}{ {\rm GeV} }
\newcommand{\TEV}{ {\rm TeV} }
\newcommand{\bea}{\begin{eqnarray}}   
\newcommand{\eea}{\end{eqnarray}}
\newcommand{\bear}{\begin{array}}  
\newcommand {\eear}{\end{array}}
\newcommand{\bef}{\begin{figure}}  
\newcommand {\eef}{\end{figure}}
\newcommand{\bec}{\begin{center}}  
\newcommand {\eec}{\end{center}}
\newcommand{\non}{\nonumber}  
\newcommand {\eqn}[1]{\beq {#1}\eeq}
\newcommand{\la}{\left\langle}  
\newcommand{\ra}{\right\rangle}
\newcommand{\ds}{\displaystyle}
\def\SEC#1{Sec.~\ref{#1}}
\def\FIG#1{Fig.~\ref{#1}}
\def\EQ#1{Eq.~(\ref{#1})}
\def\EQS#1{Eqs.~(\ref{#1})}
\def\GEV#1{10^{#1}{\rm\,GeV}}
\def\MEV#1{10^{#1}{\rm\,MeV}}
\def\KEV#1{10^{#1}{\rm\,keV}}
\def\lrf#1#2{ \left(\frac{#1}{#2}\right)}
\def\lrfp#1#2#3{ \left(\frac{#1}{#2} \right)^{#3}}

%


\preprint{UT-15-18}
\title{Higgs inflation and suppression of axion isocurvature perturbation}
\renewcommand{\thefootnote}{\alph{footnote}}

\author{
Kazunori Nakayama$^{a,b}$
and 
Masahiro Takimoto$^{a}$}

\affiliation{
 $^a$Department of Physics, University of Tokyo, Tokyo 113-0033, Japan\\
 $^b$Kavli Institute for the Physics and Mathematics of the Universe, UTIAS, University of Tokyo, Kashiwa 277-8583, Japan
  }

\begin{abstract}

We point out that cosmological constraint from the axion isocurvature perturbation is relaxed
if the Higgs field obtains a large field value during inflation in the DFSZ axion model.
This scenario is consistent with the Higgs inflation model, in which two Higgs doublets have non-minimal couplings 
and play a role of inflaton.

\end{abstract}

\maketitle


\section{Introduction}

Strong CP problem is one of the remaining mysteries of the standard model (SM).
Among solutions to the strong CP problem proposed so far, the axion solution~\cite{Peccei:1977hh} is the most attractive one~\cite{Kim:1986ax}.
In axion models, the global U(1) symmetry, which is called Peccei-Quinn (PQ) symmetry, is spontaneously broken and
the resulting pseudo Nambu-Goldstone (NG) boson takes a roll of axion to solve the strong CP problem.
However, cosmological effects of the axion are highly non-trivial. If the PQ symmetry is broken after inflation, 
topological defects are formed and they may be harmful for cosmology.
On the other hand, if the PQ symmetry is broken during inflation, the axion obtains quantum fluctuations which result in 
the cold dark matter (CDM) isocurvature perturbation~\cite{Kawasaki:2013ae}.

In the standard cosmological scenario in which the PQ symmetry is already broken during inflation and the PQ scalar $\phi$ is stabilized at the
potential minimum $|\phi|=f_a/\sqrt{2}$, with $f_a$ being the PQ scale, the magnitude of the axion isocurvature perturbation is given by
\begin{equation}
	\frac{\delta a}{a} \simeq \frac{H_{\rm inf}}{2\pi f_a \theta_i},
\end{equation}
where $H_{\rm inf}$ denotes the Hubble scale during inflation and $\theta_i$ is the initial misalignment angle.
The CDM isocurvature perturbation is given by $S_{\rm CDM}=r_a (2\delta a/a)$ with $r_a$ being the axion fraction in the present CDM energy density,
and is evaluated as
\begin{equation}
	S_{\rm CDM} \simeq 1\times 10^{-5}\theta_i \left( \frac{f_a}{10^{12}\,{\rm GeV}} \right)^{0.18}\left( \frac{H_{\rm inf}}{10^{7}\,{\rm GeV}} \right).
\end{equation}
The Planck constraint on the uncorrelated CDM isocurvature perturbation reads $S_{\rm CDM} \lesssim 1.4\times 10^{-6}$~\cite{Ade:2015lrj}.
For the axion window $10^9\,{\rm GeV}\lesssim f_a \lesssim 10^{12}\,{\rm GeV}$, the inflation scale is constrained to be
$H_{\rm inf} \lesssim 10^{7-10}\,{\rm GeV}$ taking account of non-Gaussian fluctuations~\cite{Kawasaki:2008sn,Kobayashi:2013nva}.
In particular, if we demand that the axion is a dominant component of CDM, the constraint reads $H_{\rm inf} \lesssim 10^{7}\,{\rm GeV}$.
It excludes high-scale inflation models unless PQ symmetry is restored after inflation.
But in a class of axion models with domain wall number larger than one, such as the DFSZ axion model~\cite{Dine:1981rt},
the formation of axionic domain wall leads to a cosmological disaster.

There are some ways to solve or relax the axion isocurvature problem.
For example, the PQ scale during inflation may be much larger than that in the present vacuum.
In this case, the isocurvature perturbation is suppressed by the ratio of the PQ scale at present and during inflation~\cite{Linde:1990yj}:
see also~\cite{Nakayama:2012zc,Moroi:2014mqa,Fairbairn:2014zta,Choi:2015zra}.
Another way is to make the axion heavy during inflation so that it does not obtain large scale quantum fluctuations
due to the stronger QCD~\cite{Dvali:1995ce,Choi:1996fs,Jeong:2013xta,Choi:2015zra} or the explicit PQ breaking term~\cite{Dine:2004cq,Higaki:2014ooa}.
The non-minimal kinetic term of the axion may also relax the constraint~\cite{Folkerts:2013tua}.

In this letter we consider a variant type of the scenario that suppresses the axion isocurvature perturbation.
To this end, we consider the DFSZ axion model in which there are two Higgs doublets with PQ charges.
If the Higgs bosons have large field values during inflation, the effective PQ scale is much larger than the present vacuum;
the angular component of PQ scalar is massive during inflation and does not obtain quantum fluctuations while
the massless mode mostly consists of the pseudo scalar Higgs, which later becomes massive and does not lead to observable isocurvature perturbations.
Actually, the effect of small mixing between the pseudo scalar Higgs and PQ scalar leads to the axion isocurvature perturbation,
which is suppressed by the large field value of Higgs during inflation.\footnote{
	A similar idea has been proposed very recently in Ref.~\cite{Choi:2015zra}.
}
In particular, we will show that it can be consistent with the Higgs inflation scenario~\cite{Bezrukov:2007ep}
in which the Higgs boson takes a roll of inflaton due to the non-minimal coupling to gravity.

\section{Toy model analysis}

Before going into the detailed study, we first roughly sketch the basic idea in a toy model.
Let us consider the model with two gauge singlet complex scalars $\phi$ and $S$, both have opposite PQ charges.
Then we write down the Lagrangian as
\begin{align}
	\mathcal L &= -|\partial_\mu \phi|^2 -|\partial_\mu S|^2 - V-\lambda(\phi^2S^2+{\rm h.c.}),  \label{toy}
\end{align}
where $\lambda$ can be taken real and positive without loss of generality, and
\begin{align}
	V= \lambda_\phi \left(|\phi|^2-\frac{\eta^2}{2}\right)^2+m_S^2 |S|^2 + \lambda_S|S|^4 + \lambda_{\phi S}|\phi|^2|S|^2.
\end{align}
Here $\lambda_\phi$, $\lambda_S$ and $m_S^2$ are real and positive.
$\eta$ is defined as $\eta\equiv N_{\rm DW}f_a$ with $N_{\rm DW}$ being the domain wall number which is a model dependent integer.
In the following, we consider the case $\lambda_{\phi S}<2\lambda$ for simplicity.
For the stability of the potential at large $|S|$ and $|\phi|$, we need
\begin{align}
	\lambda_\phi\lambda_S > (\lambda-\lambda_{\phi S}/2)^2.
\end{align}
A term proportional to $\phi S$ can be forbidden by a $Z_2$ symmetry, under which $S$ transforms as $S\to -S$.
The vacuum lies at $S=0$ and $|\phi|=\eta/\sqrt{2}$ where the NG mode, axion, is exactly the angular component of $\phi$:
\begin{equation}
	\phi = \frac{\eta}{\sqrt 2} \exp\left( i \frac{a_\phi}{\eta} \right).
\end{equation}

However, the role of $\phi$ and $S$ are inverted if $S$ has a large field value during inflation.
Actually it can be as large as $v_S\equiv |S| \sim H_{\rm inf}/\sqrt{\lambda_S}$
and $S$ can be the inflaton consistent with Planck observations for $\lambda_S \sim 10^{-13}$ by introducing higher order non-renormalizable terms
to make the $S$ potential polynomial~\cite{Nakayama:2013jka} or for $\lambda_S\sim \mathcal O(1)$ by adding non-minimal couplings to gravity~\cite{Bezrukov:2007ep} or non-minimal kinetic terms of $S$~\cite{Takahashi:2010ky}.
Let us write $\phi$ and $S$ as
\begin{equation}
	\phi = \frac{v_\phi}{\sqrt 2} \exp\left( i \frac{a_\phi}{v_\phi} \right),~~~S = \frac{v_S}{\sqrt 2} \exp\left( i \frac{a_S}{v_S} \right).
\end{equation}
In the vacuum, $v_\phi=\eta$ and $v_S=0$. But they need not coincide with the vacuum values in the early universe.
We assume that initially $S$ has a large field value $v_S$.
Then $v_\phi$ is given by
\begin{equation}
	v_\phi \simeq {\rm max}\left[\eta,\,\sqrt{\frac{2\lambda-\lambda_{\phi S}}{2\lambda_\phi}}v_S \right],
\end{equation}
If we introduce a Hubble induced mass term for $\phi$ as $V \sim -c^2H^2|\phi|^2$ with $c$ being a constant,
it can also stabilize the PQ scalar at around $v_\phi \sim cH/\sqrt{\lambda_\phi}$.
The massless mode, which we denote by $a$, mostly consists of the angular component of $S$ if $v_S \gg v_\phi$:
\begin{align}
	a \simeq a_{S} - \frac{v_\phi}{v_S} a_\phi.
\end{align}
It is mostly the pseudo scalar $a_S$ that obtains quantum fluctuations of $\sim H_{\rm inf}/2\pi$.
On the other hand, the orthogonal component, mostly consisting of $a_\phi$, obtains a mass of
\begin{equation}
	m^2_{a_\phi} \simeq \lambda v_S^2.
\end{equation}
Therefore, if this exceeds the Hubble scale during inflation, $a_\phi$ does not develop quantum fluctuations.
This condition is written as
\begin{equation}
	\lambda \gtrsim \left( \frac{H_{\rm inf}}{v_S} \right)^2.  \label{cond1}
\end{equation}
But $a_\phi$ constitutes a small fraction $(\sim v_\phi/v_S)$ of massless mode $a$, hence the fluctuation of $a_\phi$ 
during inflation is estimated as
\begin{equation}
	\left(\delta a_\phi\right)_{\rm inf} \sim \frac{v_\phi}{v_S} \frac{H_{\rm inf}}{2\pi}.
\end{equation}
Thus we obtain the axion isocurvature fluctuation at the QCD phase transition as
\begin{equation}
	\left(\frac{\delta a_\phi}{a_\phi} \right)_{\rm QCD} \simeq \frac{H_{\rm inf}}{2\pi v_S \theta_i}.
\end{equation}
Compared with the standard scenario with the broken PQ symmetry during inflation,
the isocurvature perturbation is suppressed by the factor $\sim f_a/v_S \ll 1$.
Since we do not introduce an explicit PQ breaking, there is always a NG mode, but the massless mode during inflation needs not coincide with
that in the present vacuum. The axion at present was massive during inflation, while the massive mode was massless.
The CDM isocurvature perturbation then is given by
\begin{equation}
	S_{\rm CDM} \simeq 1\times 10^{-5}\theta_i \left( \frac{f_a}{10^{12}\,{\rm GeV}} \right)^{0.18}\left( \frac{H_{\rm inf}}{10^{7}\,{\rm GeV}} \right)
	\left( \frac{f_a}{v_S} \right).
	\label{SCDM}
\end{equation}
Thus inflation scale as large as $H_{\rm inf}\sim 10^{13}$\,GeV may be allowed for $v_S \sim 10^{18}$\,GeV
even if the axion is a dominant component of CDM.

In this class of scenario, $S$ has a large field value initially and hence it begins a coherent oscillation after inflation and 
must decay into SM particles.
Later we will see that $S$ itself can be the inflaton.
The oscillation of $S$ potentially causes resonant particle production of axions~\cite{Kofman:1997yn} and
may induce the dynamical motion of $\phi$.
We denote the frequency of $S$ after the onset of the oscillation as $m_S^{\rm osc}$.
%
%
First, we consider the axion production due to the oscillation of $S$.
If the condition
\begin{align}
\label{cond3}
	\frac{\sqrt{\lambda}v_S}{m_S^{\rm osc}}<1,
\end{align}
holds, the axion production of the broad resonance type does not occur even just after the onset of the oscillation of $S$.
In such a case, the narrow resonance would harmfully produce axion particles. 
However, the narrow resonance is ineffective if the decay rate of $S$ is sizable~\cite{Ema:2015oaa}.
Second, let us consider the motion of $\phi$ caused by the oscillation of $S$.
The potential of $\phi$ varies as $S$ oscillates and $\phi$ may also oscillate around the origin.
In such a case, the nonthermal restoration of the PQ symmetry~\cite{Kofman:1995fi,Kasuya:1996ns,Moroi:2013tea,Kawasaki:2013iha} 
could occur due to the parametric resonance at the origin because of the self interaction of $\phi$. 
The parametric resonance at the origin becomes most effective when $S$ decays just after the onset of oscillation of $S$. 
In such a case, the condition for the PQ symmetry not to be restored is obtained from the condition that
the number of $\phi$ oscillation within one Hubble time remains less than $\sim \mathcal O(10^2)$ until the oscillation amplitude of
$\phi$ reduces to $\sim f_a$~\cite{Kawasaki:2013iha}. This leads to a constraint
\begin{equation}
	\left(\frac{H_{\rm inf}}{\sqrt{\lambda_\phi} f_a}\right)^n \lesssim \mathcal O (10^2).
	\label{cond4}
\end{equation}
where $n=1/2\,(1)$ for the matter (radiation) dominated background evolution, where we have assumed
the inequality (\ref{cond1}) is marginally satisfied: $\sqrt{\lambda} v_S \sim H_{\rm inf}$.

Next, let us consider the thermal effects on $\phi$.
In order for the PQ symmetry not to be restored after inflation due to thermal effects, we need
\begin{equation}
	\lambda_{\phi S} \lesssim \lambda_{\phi} \left( \frac{f_a}{T_{\rm max}} \right)^2,   \label{cond2}
\end{equation}
where $T_{\rm max}$ denotes the maximum temperature after inflation and assumed that $S$ is in thermal equilibrium
due to interactions with SM particles.
Otherwise, the PQ symmetry may be restored after inflation and isocurvature constraint disappears, 
but there arises a serious domain wall problem.
Conditions (\ref{cond1}), (\ref{cond3}), (\ref{cond4}) and (\ref{cond2}) are satisfied for reasonable choice of parameters.

\section{DFSZ axion model}

Let us move to the DFSZ axion model, two Higgs doublets are introduced~\cite{Branco:2011iw}.
In this model, $S^2$ in the toy model (\ref{toy}) may be identified with the 
gauge invariant combination of the Higgs doublets $H_u H_d$. The action in the Jordan frame is given by
\begin{align}
	S = \int d^4x \sqrt{-g_J}\left(\mathcal L_{g} +\mathcal L_J -V_J + \mathcal L^{\rm (SM)}_J\right),
\end{align}
where
\begin{align}
	\mathcal L_{g} &=  \left(\frac{M_P^2}{2} + \xi_u |H_u|^2+\xi_d|H_d|^2\right) R_J,\\
	\mathcal L_{J} &= -|\mathcal D_\mu H_u|^2 -  |\mathcal D_\mu H_d|^2 -  |\partial_\mu \phi|^2,\\ \nonumber
	V_J &= m_u^2|H_u|^2 +  m_d^2|H_d|^2 + (\lambda \phi^2 H_u H_d + {\rm h.c.}) \\ \nonumber
	    &+ \lambda_u |H_u|^4 +\lambda_d |H_d|^4 \\ \nonumber
	    &+ \lambda_{ud} |H_u|^2 |H_d|^2 + \lambda_{ud}' |H_u H_d|^2 \\
	    &+ \lambda_{u\phi}|H_u|^2 |\phi|^2 + \lambda_{d\phi}|H_d|^2 |\phi|^2 + V(|\phi|),\\
	 V(|\phi|) &= \lambda_\phi\left(|\phi|^2 - \frac{\eta^2}{2}\right)^2,
\end{align}
and $\mathcal L^{\rm (SM)}$ contains kinetic terms for SM quarks, leptons and gauge bosons and yukawa couplings between
quarks/leptons and Higgs bosons and $R_J$ is the Ricci scalar.
Quantities with subscript $J$ indicate those in the Jordan frame.
Here $H_u$ is assumed to couple with up-type quarks and $H_d$ with down-type quarks and charged leptons.
This can be done by assigning the PQ charge, e.g., $1$ to the PQ scalar $\phi$, $-2$ to $H_u$ and $2$ to right-handed up-type quarks.
In this model, the domain wall number $N_{\rm DW}$ is equal to $6$.
In the present vacuum, the PQ scalar obtains a VEV of $|\phi|=\eta/\sqrt{2}$ and Higgs bosons have electroweak scale VEVs.
Note that since the Higgs bosons also have PQ charges, the axion is a mixture of the angular components of $\phi$ and Higgs bosons:
\begin{align}
	a \simeq a_\phi - \frac{\sin 2\beta}{\eta}\left( v_d a_u + v_u a_d\right),
\end{align}
where we have defined
\begin{equation}
	H_u^0 = \frac{v_u}{\sqrt 2} \exp\left( i \frac{a_u}{v_u} \right),~~H_d^0 = \frac{v_d}{\sqrt 2} \exp\left( i \frac{a_d}{v_d} \right),
\end{equation}
and $\tan\beta\equiv v_u/v_d$.
However, the mixing angle is suppressed by the ratio $v_{u,d}/f_a$ and hence the massless mode almost consists of $a_\phi$.
Actually it is this small mixing that admits the axion-gluon-gluon coupling, required for solving the strong CP problem.
Another massless mode, $a_G=\cos\beta a_d-\sin\beta a_u$, is eaten by the $Z$-boson and 
the other orthogonal combination becomes massive, 
which we identify as the pseudo scalar Higgs and denote by $a_h$.
For large enough $v_u, v_d \gg v_\phi$, however, the axion is identified as
\begin{equation}
	a \simeq \cos\beta a_u + \sin\beta a_d - \frac{v_\phi}{v\sin2\beta} a_\phi,
\end{equation}
with $v\equiv \sqrt{v_u^2+v_d^2}$.
Actually, in the Higgs inflation model, $v_u$ and $v_d$ can take so large values that the axion isocurvature perturbation is suppressed
as shown in the toy model in the previous section.

The detailed analysis of Higgs inflation with two Higgs doublets are found in Ref.~\cite{Gong:2012ri}.
Thanks to the PQ symmetry, allowed terms are limited.
After the conformal transformation $g^E_{\mu\nu} = \Omega^2 g_{\mu\nu}^J$ where
\begin{equation}
	\Omega^2 = 1+\frac{2\xi_u|H_u|^2}{M_P^2} + \frac{2\xi_d |H_d|^2}{M_P^2},
\end{equation}
the Einstein frame action becomes
\begin{equation}
	S = \int d^4x \sqrt{-g_E}\left( \frac{M_P^2}{2}R_E + \mathcal L_E -V_E +\mathcal L^{\rm (SM)}_E \right),
\end{equation}
with subscript $E$ indicating the Einstein frame.
The scalar potential is given by
\begin{equation}
	V_E(H_u, H_d,\phi) = \frac{V_J}{\Omega^4}.
\end{equation}
Focusing on only terms with fourth powers of $h_u$ and $h_d$, it is expressed as
\begin{equation}
	\frac{V_E}{M_P^4} = \frac{\lambda_u v_u^4 + \lambda_d v_d^4 + \bar\lambda_{ud}v_u^2v_d^2 }{4(1+\xi_u v_u^2 + \xi_d v_d^2)^2}.
\end{equation}
where $\bar\lambda_{ud} \equiv \lambda_{ud}+\lambda_{ud}'$.
Thus it is easy to see that the potential becomes flat for $v_u\gg M_P/\sqrt{\xi_u}$ or $v_d\gg M_P/\sqrt{\xi_d}$.
Terms involving $\phi$ do not affect the inflaton dynamics as we will see.
There is a stable inflationary path along
\begin{equation}
	\frac{v_u^2}{v_d^2}=\frac{2\lambda_d\xi_u - \bar\lambda_{ud}\xi_d}{2\lambda_u\xi_d - \bar\lambda_{ud}\xi_u},
\end{equation}
if $2\lambda_d\xi_u - \bar\lambda_{ud}\xi_d > 0$ and $2\lambda_u\xi_d - \bar\lambda_{ud}\xi_u > 0$.
Then the potential energy for the inflaton is given by
\begin{equation}
	\frac{V_E}{M_P^4} = \frac{\lambda_u\lambda_d- \bar\lambda_{ud}^2/4 }{4(\lambda_u\xi_d^2 + \lambda_d\xi_u^2 - \bar\lambda_{ud}\xi_u\xi_d)}
	\left(1-e^{-2\chi/\sqrt{6}M_P} \right)^2,
\end{equation}
where $\chi$ is the canonically normalized field in the large field limit:
\begin{equation}
	\chi = \sqrt{\frac{3}{2}}M_P \log \left(1+\frac{\xi_uv_u^2}{M_P^2} + \frac{\xi_d v_d^2}{M_P^2} \right).
\end{equation}
Thus it reduces to the single field Higgs inflation model.
Assuming $\lambda_u\sim \lambda_d\sim \lambda_{ud} \sim \lambda_{ud}'$ and $\xi_u \sim \xi_d$, we need
\begin{equation}
	\xi \sim 5\times 10^4 \sqrt{\lambda_{\rm eff}},
\end{equation}
to reproduce the density perturbation observed by Planck~\cite{Ade:2015lrj}, 
where $\xi$ and $\lambda_{\rm eff}$ denote typical values of $\xi$'s and $\lambda$'s, respectively.
To explain the 125\,GeV Higgs boson, $\lambda_{\rm eff}$ should be $\mathcal O(1)$,
but the running tends to make $\lambda$ smaller at high energy scale and 
hence $\xi$ may be able to take a smaller value~\cite{Salvio:2013rja,Hamada:2014wna}.

Next, let us see the behavior of the PQ scalar during inflation. It obtains a VEV of
\begin{align}
	v_\phi \simeq {\rm max}\left[ \eta, ~\sqrt{\frac{\lambda}{\lambda_\phi}}v_{h}  \right],
\end{align}
where $v_h(\sim v_u \sim v_d\sim 10M_P/\sqrt{\xi})$ denotes the typical field value of the Higgs boson during inflation.
The dynamics after inflation is effectively the same as that studied in the previous section.
Note that $a_\phi$ obtains a mass of
\begin{align}
	m_{a_\phi}^2\simeq \frac{\lambda v_h^2}{\Omega^2}\simeq \frac{\lambda M_P^2}{\xi}.
\end{align}
Thus we impose the condition
\begin{align}
	\lambda \gtrsim \frac{\xi H_{\rm inf}^2}{M_P^2},
\end{align}
to suppress the fluctuation of $a_\phi$ during inflation.
Also note that the reheating temperature after inflation may be as high as $\sim 10^{13}\,$GeV taking account of
efficient particle production and thermal dissipative effects~\cite{Bezrukov:2008ut,GarciaBellido:2008ab,Mukaida:2012qn}.
To avoid the PQ symmetry restoration, we need a condition similar to (\ref{cond2}):
\begin{equation}
	\lambda_{u\phi}, \lambda_{d\phi} \lesssim \lambda_{\phi} \left( \frac{f_a}{T_{\rm max}} \right)^2.
\end{equation}

The pseudo scalar Higgs, $a_h$, is almost massless during inflation and develops quantum fluctuations.
Note that the canonically normarized field is $\tilde a_h \equiv a_h M_P/\sqrt{\xi}v_h$ which obtains a fluctuation of $\delta \tilde a_h \simeq H_{\rm inf}/2\pi$,
while the mixing between $a_\phi$ and $a_h$ is given by $\sim v_\phi/v_h$.
Then we find the effective fluctuation of $a_\phi$ as $\delta a_\phi/v_\phi \sim \sqrt{\xi} H_{\rm inf}/(2\pi M_P).$
Thus the axion isocurvature fluctuation is given by
\begin{equation}
	\left(\frac{\delta a_\phi}{a_\phi}\right)_{\rm QCD} \sim \frac{\sqrt{\xi} H_{\rm inf}}{2\pi M_P \theta_i}.  \label{axioniso}
\end{equation}
The CDM isocurvature perturbation is given by (\ref{SCDM}) with $v_S \to M_P/\sqrt{\xi}$.
Thus the constraint is significantly relaxed.

\section{Discussion}

In this letter we have shown a way to suppress the axion isocurvature perturbation.
Even if the PQ scalar itself does not have a field value much larger than $f_a$ during inflation,
other PQ charged scalar fields can have large field values and it results in suppression of the isocurvature perturbation.
Interestingly, the electroweak Higgs bosons may take such roles in the DFSZ axion model,
and the Higgs bosons can also be the inflaton.

So far, we have considered the quartic interaction term between the PQ scalar and the Higgs bosons $\lambda \phi^2 H_u H_d +{\rm h.c.}$.
There are other possible choices. For example, let us consider the case with the cubic term $V = \mu\phi H_u H_d +{\rm h.c.}$.
In this case, $\phi$ settles around $\phi\sim \mu/\lambda_{u\phi,d\phi}$ 
and has the mass of $\sim \lambda_{u\phi,d\phi}v_h^2$ during the inflation. After the inflation, $\phi$ starts to oscillate.
If $\mu\lesssim f_a$ holds, $S$ will not pass
through the origin and the restoration of the PQ symmetry due to the particle production does not occur. 
The higher dimensional terms may also play important roles such as stabilization of the PQ scalar at some field value.

Although we considered the case of Higgs boson as the inflaton, we can also regard the PQ scalar as the inflaton in a similar way,
by introducing a non-minimal coupling as~\cite{Fairbairn:2014zta}\footnote{
	The PQ scalar as the inflaton in the context of running kinetic inflation was mentioned in Ref.~\cite{Higaki:2014ooa}.
}
\begin{eqnarray}
	\mathcal L = \xi_\phi |\phi|^2 R_J.
\end{eqnarray}
The PQ scalar can take a role of inflaton for $\xi_\phi \sim 5\times 10^4\sqrt{\lambda_\phi}$.
By noting that the canonically normalized axion field at $v_\phi > M_P/\sqrt{\xi_\phi}$ is given by
$\tilde a_\phi \equiv a_\phi M_P/\sqrt{\xi_\phi} v_\phi$ and it obtains quantum fluctuations of $\simeq H_{\rm inf}/2\pi$,
we would have the axion isocurvature perturbation of the same expression as (\ref{axioniso}),
if the PQ symmetry is never restored thereafter.
There is a subtlety, however, in this scenario.
Because the canonically normalized PQ scalar has an oscillation amplitude of $\sim M_P$ just after inflation,
it induces efficient axion production when it passes through the origin of the potential,
and the PQ symmetry may be restored nonthermally, leading to formation of axion domain walls
if the domain wall number is larger than one~\cite{Kawasaki:2013iha}.
We can arrange the model so that the Higgs bosons also have large VEVs during and after inflation 
just as we have done in the previous section for $\phi$.
Then the PQ symmetry is broken by Higgs VEVs even though $\phi$ is nonthermally trapped around the origin.
Even in such a case, however, the model has a $Z_2$ symmetry under which $\phi\to -\phi$ and anything else uncharged,
hence the domain wall formation is not avoided when $\phi$ relaxes to the minimum.
Again, introducing $V = \mu\phi H_u H_d +{\rm h.c.}$ in the potential, instead of $\lambda \phi^2 H_u H_d +{\rm h.c.}$,
by assigning appropriate PQ charges, may help the situation but a complete analysis is beyond the scope of this paper.

\section*{Acknowledgments}

This work was supported by Grant-in-Aid for Scientific research 26104009 (KN), 
26247042 (KN), 26800121 (KN), MEXT, Japan.
The work of M.T. are supported in part by JSPS Research Fellowships
for Young Scientists and by the Program for Leading Graduate Schools, MEXT, Japan.




\begin{thebibliography}{99}


\bibitem{Peccei:1977hh} 
  R.~D.~Peccei and H.~R.~Quinn,
  Phys.\ Rev.\ Lett.\  {\bf 38}, 1440 (1977).
  
\bibitem{Kim:1986ax} 
  For a review, see J.~E.~Kim,
  Phys.\ Rept.\  {\bf 150}, 1 (1987);
  J.~E.~Kim and G.~Carosi,
  Rev.\ Mod.\ Phys.\  {\bf 82}, 557 (2010)
  [arXiv:0807.3125 [hep-ph]].
  
\bibitem{Kawasaki:2013ae} 
  For a review on axion cosmology, see M.~Kawasaki and K.~Nakayama,
  Ann.\ Rev.\ Nucl.\ Part.\ Sci.\  {\bf 63}, 69 (2013)
  [arXiv:1301.1123 [hep-ph]].
  
\bibitem{Ade:2015lrj} 
  P.~A.~R.~Ade {\it et al.}  [Planck Collaboration],
  arXiv:1502.02114 [astro-ph.CO].
  
\bibitem{Kawasaki:2008sn} 
  M.~Kawasaki, K.~Nakayama, T.~Sekiguchi, T.~Suyama and F.~Takahashi,
  JCAP {\bf 0811}, 019 (2008)
  [arXiv:0808.0009 [astro-ph]];
  C.~Hikage, M.~Kawasaki, T.~Sekiguchi and T.~Takahashi,
  JCAP {\bf 1307}, 007 (2013)
  [arXiv:1211.1095 [astro-ph.CO]].
  
\bibitem{Kobayashi:2013nva} 
  T.~Kobayashi, R.~Kurematsu and F.~Takahashi,
  JCAP {\bf 1309}, 032 (2013)
  [arXiv:1304.0922 [hep-ph]].
  
  
  
\bibitem{Dine:1981rt}
  M.~Dine, W.~Fischler and M.~Srednicki,
  Phys.\ Lett.\  B {\bf 104}, 199 (1981);
  A.~R.~Zhitnitsky,
  Sov.\ J.\ Nucl.\ Phys.\  {\bf 31}, 260 (1980)
  [Yad.\ Fiz.\  {\bf 31}, 497 (1980)].
  

\bibitem{Linde:1990yj} 
  A.~D.~Linde and D.~H.~Lyth,
  Phys.\ Lett.\ B {\bf 246}, 353 (1990);
  A.~D.~Linde,
  Phys.\ Lett.\ B {\bf 259}, 38 (1991).
  
\bibitem{Nakayama:2012zc} 
  K.~Nakayama and N.~Yokozaki,
  JHEP {\bf 1211}, 158 (2012)
  [arXiv:1204.5420 [hep-ph]].
  
\bibitem{Moroi:2014mqa} 
  T.~Moroi, K.~Mukaida, K.~Nakayama and M.~Takimoto,
  JHEP {\bf 1411}, 151 (2014)
  [arXiv:1407.7465 [hep-ph]].
  
\bibitem{Fairbairn:2014zta} 
  M.~Fairbairn, R.~Hogan and D.~J.~E.~Marsh,
  Phys.\ Rev.\ D {\bf 91}, no. 2, 023509 (2015)
  [arXiv:1410.1752 [hep-ph]].

\bibitem{Choi:2015zra} 
  K.~Choi, E.~J.~Chun, S.~H.~Im and K.~S.~Jeong,
  arXiv:1505.00306 [hep-ph].
  
\bibitem{Dvali:1995ce} 
  G.~R.~Dvali,
  hep-ph/9505253.
  
\bibitem{Choi:1996fs} 
  K.~Choi, H.~B.~Kim and J.~E.~Kim,
  Nucl.\ Phys.\ B {\bf 490}, 349 (1997)
  [hep-ph/9606372].
  
\bibitem{Jeong:2013xta} 
  K.~S.~Jeong and F.~Takahashi,
  Phys.\ Lett.\ B {\bf 727}, 448 (2013)
  [arXiv:1304.8131 [hep-ph]].
  
\bibitem{Dine:2004cq} 
  M.~Dine and A.~Anisimov,
  JCAP {\bf 0507}, 009 (2005)
  [hep-ph/0405256].
  
\bibitem{Higaki:2014ooa} 
  T.~Higaki, K.~S.~Jeong and F.~Takahashi,
  Phys.\ Lett.\ B {\bf 734}, 21 (2014)
  [arXiv:1403.4186 [hep-ph]].
  
\bibitem{Folkerts:2013tua} 
  S.~Folkerts, C.~Germani and J.~Redondo,
  Phys.\ Lett.\ B {\bf 728}, 532 (2014)
  [arXiv:1304.7270 [hep-ph]].
  
\bibitem{Bezrukov:2007ep} 
  F.~L.~Bezrukov and M.~Shaposhnikov,
  Phys.\ Lett.\ B {\bf 659}, 703 (2008)
  [arXiv:0710.3755 [hep-th]].
  
\bibitem{Nakayama:2013jka} 
  K.~Nakayama, F.~Takahashi and T.~T.~Yanagida,
  Phys.\ Lett.\ B {\bf 725}, 111 (2013)
  [arXiv:1303.7315 [hep-ph]];
  JCAP {\bf 1308}, 038 (2013)
  [arXiv:1305.5099, arXiv:1305.5099 [hep-ph]].
 
\bibitem{Takahashi:2010ky} 
  F.~Takahashi,
  Phys.\ Lett.\ B {\bf 693}, 140 (2010)
  [arXiv:1006.2801 [hep-ph]];
  K.~Nakayama and F.~Takahashi,
  JCAP {\bf 1011}, 009 (2010)
  [arXiv:1008.2956 [hep-ph]].
    
  %
  %
  %
\bibitem{Kofman:1997yn} 
  L.~Kofman, A.~D.~Linde and A.~A.~Starobinsky,
  Phys.\ Rev.\ D {\bf 56}, 3258 (1997)
  [hep-ph/9704452].
  
\bibitem{Ema:2015oaa} 
  Y.~Ema, R.~Jinno, K.~Mukaida and K.~Nakayama,
  arXiv:1504.07119 [gr-qc].
  
\bibitem{Kofman:1995fi} 
  L.~Kofman, A.~D.~Linde and A.~A.~Starobinsky,
  Phys.\ Rev.\ Lett.\  {\bf 76}, 1011 (1996)
  [hep-th/9510119].
  
\bibitem{Kasuya:1996ns} 
  S.~Kasuya, M.~Kawasaki and T.~Yanagida,
  Phys.\ Lett.\ B {\bf 409}, 94 (1997)
  [hep-ph/9608405];
  Phys.\ Lett.\ B {\bf 415}, 117 (1997)
  [hep-ph/9709202].
  
\bibitem{Moroi:2013tea} 
  T.~Moroi, K.~Mukaida, K.~Nakayama and M.~Takimoto,
  JHEP {\bf 1306}, 040 (2013)
  [arXiv:1304.6597 [hep-ph]].
  
\bibitem{Kawasaki:2013iha} 
  M.~Kawasaki, T.~T.~Yanagida and K.~Yoshino,
  JCAP {\bf 1311}, 030 (2013)
  [arXiv:1305.5338 [hep-ph]].

  
\bibitem{Branco:2011iw} 
  For a review on two Higgs doublet models, see G.~C.~Branco, P.~M.~Ferreira, L.~Lavoura, M.~N.~Rebelo, M.~Sher and J.~P.~Silva,
  Phys.\ Rept.\  {\bf 516}, 1 (2012)
  [arXiv:1106.0034 [hep-ph]].
  
\bibitem{Gong:2012ri} 
  J.~O.~Gong, H.~M.~Lee and S.~K.~Kang,
  JHEP {\bf 1204}, 128 (2012)
  [arXiv:1202.0288 [hep-ph]].
  
\bibitem{Salvio:2013rja} 
  A.~Salvio,
  Phys.\ Lett.\ B {\bf 727}, 234 (2013)
  [arXiv:1308.2244 [hep-ph]].
  
\bibitem{Hamada:2014wna} 
  Y.~Hamada, H.~Kawai, K.~y.~Oda and S.~C.~Park,
  Phys.\ Rev.\ D {\bf 91}, no. 5, 053008 (2015)
  [arXiv:1408.4864 [hep-ph]].
  
\bibitem{Bezrukov:2008ut} 
  F.~Bezrukov, D.~Gorbunov and M.~Shaposhnikov,
  JCAP {\bf 0906}, 029 (2009)
  [arXiv:0812.3622 [hep-ph]].
  
\bibitem{GarciaBellido:2008ab} 
  J.~Garcia-Bellido, D.~G.~Figueroa and J.~Rubio,
  Phys.\ Rev.\ D {\bf 79}, 063531 (2009)
  [arXiv:0812.4624 [hep-ph]].
  
\bibitem{Mukaida:2012qn} 
  K.~Mukaida and K.~Nakayama,
  JCAP {\bf 1301}, 017 (2013)
  [arXiv:1208.3399 [hep-ph]];
  JCAP {\bf 1303}, 002 (2013)
  [arXiv:1212.4985 [hep-ph]];
  K.~Mukaida, K.~Nakayama and M.~Takimoto,
  JHEP {\bf 1312}, 053 (2013)
  [arXiv:1308.4394 [hep-ph]].
    
  
\end{thebibliography}
\end{document}